\documentclass[12pt]{iopart}

\usepackage{graphicx}
\usepackage{color}

\def\spose#1{\hbox to 0pt{#1\hss}}
\def\simlt{\mathrel{\spose{\lower 3pt\hbox{$\mathchar"218$}}
     \raise 2.0pt\hbox{$\mathchar"13C$}}}
\def\simgt{\mathrel{\spose{\lower 3pt\hbox{$\mathchar"218$}}
     \raise 2.0pt\hbox{$\mathchar"13E$}}}
\begin{document}


\title[A Reply to a Comment on 
``Dark matter: A phenomenological existence proof'']{A Reply to a Comment on
``Dark matter: A phenomenological existence proof''}

	\author{\textcolor{blue}{D  V  Ahluwalia-Khalilova}}

	\address{ASGBG, Apartado Postal C-600, Department of Mathematics\\
   	University of Zacatecas (UAZ), Zacatecas 98060, Mexico}

	\ead{ahluwalia@heritage.reduaz.mx}
	     

\thispagestyle{empty}

\begin{abstract}
In astro-ph/0601489,  within the framework of the 
Einsteinian general relativity,
we  made  the observation that if the universe is
described by a
spatially flat Friedmann-Robertson-Walker (FRW)
cosmology with Einsteinian cosmological
constant
then the resulting cosmology predicts a 
significant dark matter component in the 
universe. Furthermore, the phenomenologically motivated existence 
proof refrained from invoking the
 data on galactic rotational curves and gravitational lensing,
but used as input the age of the universe as deciphered from the
studies on globular clusters. 
This claim has been challenged in 
astro-ph/0603213. Here we show that the raised objection
is invalid.  It, at best, 
constitutes a trivial consistency check. As such,
we stand by our analysis, and by our conclusions, 
without reservations.

\end{abstract}
\pacs{98.80.Bp, 98.80.Jk}

\textcolor{blue}{\hrule}


\newpage

Referring the reader  to Ref. \cite{Ahluwalia-Khalilova:2006ez}
for notational details we recall that there 
we considered a FRW cosmology defined by the set

\begin{equation}
\{k=0,\; w^\Lambda=-1,\; \rho=\rho_\mathrm{m},
\;p=p_\mathrm{m}=0,\;\;\rho^\Lambda={\mathrm{constant}}\}\label{eq:set}
\end{equation}
and proved, without further assumptions, that 

\begin{equation}
\Omega_\mathrm{m}(t) : \Omega_\Lambda(t) = 1:\zeta(t)\label{eq:zeta}
\end{equation}
where
$
\zeta(t):= \sinh^2
\left(\sqrt{3 }\, t/ (2 \tau_\Lambda) \right)$, and
we defined  $\tau_\Lambda:=\sqrt{1/\Lambda}$. We took the observational
value of $\Lambda$, defined a reference unit of time, 
and explicitly stated it \textemdash~something which
the author of Ref.~\cite{Vishwakarma:2006ju} seems to have ignored in making his
objections.

To avoid semantical misinterpretation we note that we consider a 
cosmology with Einsteinian cosmological constant. To say that 
this cosmology, in any sense, does not invoke a specific value of 
$\Lambda$ carries no content (and is wrong).  This statement holds
true in the numerical sense, rather than analytical. For the latter,
no specific value of $\Lambda$ need be invoked.

Now $\zeta(t):= \Omega_\Lambda/\Omega_\mathrm{m}$ is a \textit{unique} function of $t$
\textemdash~or, if one wishes to be pedantic, of $t/\tau_\Lambda$.

Once the age of the universe is specified by some 
\textit{independent} observations,\footnote{We shall not take issue
even if this assertion was to be totally abandoned. To do so would be
against the spirit and content of our argument contained in
 Ref.~\cite{Ahluwalia-Khalilova:2006ez}.}
the considered cosmology uniquely determines
the ratio $\Omega_\mathrm{m}(t) : \Omega_\Lambda(t)$.
A graphical representation of $\zeta(t)$ is given in
\cite[Fig. 1]{Ahluwalia-Khalilova:2006ez}.
In addition, the fractional matter density is \textit{predicted} 

\begin{equation}
\Omega_\mathrm{m}(t)= (1+\zeta(t))^{-1}\,.
\end{equation}
This is the combined result of equation (\ref{eq:zeta}) and 
of fact that we are considering a spatially flat
cosmology, which requires $1 = \Omega_\mathrm{m} + \Omega_\mathrm{\Lambda}$.
A graphical representation of $\Omega_\mathrm{m}(t)$ is given in
\cite[Fig. 2]{Ahluwalia-Khalilova:2006ez}.

In Ref.~\cite{Ahluwalia-Khalilova:2006ez} we emphasized that 
this circumstance arises due to a specific
nonlinear aspect of Einstein's field equations for the 
considered FRW cosmology. Author of Ref.~\cite{Vishwakarma:2006ju}
has failed to appreciate its impact and importance in the argument
we presented.

We chose for $t$ the age of the universe as deciphered 
from the
age of the globular clusters and arrived at the claims contained
in Ref.~\cite{Ahluwalia-Khalilova:2006ez}. Briefly, we fixed a range of $t$
as indicated; that yielded a range of $\zeta$ \textit{and} $\Omega_\mathrm{m}$
(a fact completely missed by the author of  Ref.~\cite{Vishwakarma:2006ju}). 
The latter 
$\Omega$
 is the
sum of all non-relativistic matter components. Taking the standard model
contribution to $\Omega_\mathrm{m}$ as $0.05$, we deciphered that there must
exist an additional (dark matter, by definition) contribution to $\Omega_\mathrm{m}
$ in the rough range  $0.14 \le \Omega_\mathrm{dm} \le 0.30$.

We did not invoke as input any value of $\zeta$, contrary to the 
assertion of  Ref.~\cite{Vishwakarma:2006ju}. Once the range of
$t$ was fixed, it \textit{automatically} constrained $\zeta$ to
$1.9 \le \zeta \le 4.3$; that is, the ratio $\Omega_\Lambda/\Omega_\mathrm{m}$.

Finally, in order to avoid confusion we note
\begin{enumerate}

\item[$\bullet$]
We have no disagreement with the remarks made by the author of 
Ref.~\cite{Vishwakarma:2006ju} regarding the age of the universe
and the age of the globular clusters. We did not use the age of the
globular clusters but the age of the universe as deciphered from 
that age. However, even if one equated the two no significantly
different conclusion is arrived at (as long as one does not
invoke `alternate cosmologies').

\item[$\bullet$]
The quoted value of $\Omega_\Lambda = 
0.73$ invoked in Ref.~\cite{Vishwakarma:2006ju} was \textit{not} 
assumed \textit{a priori} (by us), 
but it follows once the age of the universe is specified. 
This is manifest from a glance at Fig. 1
of Ref.~\cite{Ahluwalia-Khalilova:2006ez}. However, if one chooses 
to specify the age of universe using a specific value of $\Lambda$,
or by some other means,
one obtains a simple consistency check; and indeed that  
in essence is the core result of Ref.~\cite{Vishwakarma:2006ju}.
To be precise, all interpretations of the age of the universe are
specific to cosmological models. But such a discussion, while
of much physical significance, is not the primary task of the
discussion at hand.

\item[$\bullet$]
In general the $\zeta(t)$, i.e. the ratio $\Omega_\Lambda/\Omega_\mathrm{m}$,
is not fixed for a given epoch within the context of general relativistic
cosmological models. What usually happens is that \textit{given} \textit{a priori} 
specific initial value for 
$\Omega_\Lambda/\Omega_\mathrm{m}$ its temporal evolution is predicted. For the 
model defined by (\ref{eq:set}) the situation is a bit more 
subtle~\cite{Ahluwalia-Khalilova:2006ez} due to
a nonlinear aspect of Einstein's field equations for the 
considered FRW cosmology. This is fully explained in 
Ref.~\cite{Ahluwalia-Khalilova:2006ez} and its full appreciation is necessary
to reach the results contained therein.

\end{enumerate}

As such we stand by our conclusion that
if the universe is
described by a
spatially flat Friedmann-Robertson-Walker (FRW)
cosmology with Einsteinian cosmological
constant
then the resulting cosmology predicts a 
significant dark matter component in the 
universe. Furthermore, the phenomenologically motivated existence 
proof refrains from invoking the
data on galactic rotational curves and gravitational lensing,
but uses as input (a) the age of the universe as deciphered from
studies on globular clusters, and (b) $\Omega_\mathrm{sm} \approx 0.05$,
where the subscript stands for `standard model' component.

\textcolor{red}{\hrule}

\end{document}